\documentclass[%
aip,amsmath,amssymb, reprint, linenumbers
]{revtex4-1}

\usepackage{graphicx}
\usepackage{dcolumn}
\usepackage{bm}
\usepackage[ruled,vlined]{algorithm2e}
\usepackage[utf8]{inputenc}
\usepackage[T1]{fontenc}
\usepackage{mathptmx}
\usepackage{listings}
\usepackage{tikz}
\usetikzlibrary{arrows}
\usetikzlibrary{trees}
\usepackage{midfloat}
\usepackage{cancel}
\usetikzlibrary{decorations.pathmorphing}
\usetikzlibrary{decorations.markings}
\usetikzlibrary{automata,positioning}
\usepackage{float}
\usepackage{caption}
\usepackage{simplewick}
\usepackage{csquotes}
\usepackage{graphicx}
\usepackage{subcaption}
\usepackage{multirow}
\usepackage{cleveref}
\usepackage{amsmath}
\usepackage{tabularx}
\usepackage{natbib}
\usepackage{braket}
\usepackage{todonotes}

\usepackage{xcolor}
\definecolor{codegreen}{rgb}{0,0.6,0}
\definecolor{codegray}{rgb}{0.5,0.5,0.5}
\definecolor{codepurple}{rgb}{0.58,0,0.82}
\definecolor{backcolour}{rgb}{0.95,0.95,0.92}
\lstdefinestyle{mystyle}{
    backgroundcolor=\color{backcolour},   
    commentstyle=\color{codegreen},
    keywordstyle=\color{magenta},
    numberstyle=\tiny\color{codegray},
    stringstyle=\color{codepurple},
    basicstyle=\ttfamily\footnotesize,
    breakatwhitespace=false,         
    breaklines=true,                 
    captionpos=b,                    
    keepspaces=true,                 
    numbers=left,                    
    numbersep=5pt,                  
    showspaces=false,                
    showstringspaces=false,
    showtabs=false,                  
    tabsize=2
}

\begin{document}
\nolinenumbers
\title{On-the-fly Tailoring towards a Rational Ansatz Design for Digital Quantum Simulations}
\author{Dibyendu Mondal, Sonaldeep Halder, Dipanjali Halder, Rahul Maitra}\email{rmaitra@chem.iitb.ac.in}\affiliation{Department of Chemistry, \\ Indian Institute of Technology Bombay, \\ Powai, Mumbai 400076, India} 
\date{\today}
\begin{abstract}
Recent advancements in quantum information and quantum
technology has stimulated a good deal of interest in the
development of quantum algorithms for energetics and
properties of many-fermionic systems. While the variational
quantum eigensolver is the most optimal algorithm in the
Noisy Intermediate Scale Quantum era, it is imperative 
to develop low depth quantum circuits that are physically
realizable in quantum devices. Within the
unitary coupled cluster framework, we develop COMPASS, a 
disentangled ansatz construction protocol that can 
dynamically tailor an optimal ansatz using the one 
and two-body cluster operators and a selection of 
rank-two scatterers. The construction of the ansatz may
potentially be performed in parallel quantum architecture 
through energy sorting and operator commutativity
prescreening. With 
significant reduction in the circuit depth towards 
the simulation of molecular strong correlation, COMPASS 
is shown to be highly accurate and resilient to the 
noisy circumstances of the near-term quantum hardware. 
\end{abstract}

\maketitle
\section{Introduction}
The determination of atomic and molecular energetics 
and properties is one of the most anticipated applications
of quantum computing in near term quantum computers. With
the advent of coherent multiple qubit quantum processors,
the problem with the exponential growth of the Hilbert
space that one often encounters in many-body physics and 
chemistry can be handled in a tractable manner. 
While the limited coherence time and poor gate fidelity 
are major impediments towards the implementation of
quantum many-body methods in Noisy 
Intermediate Scale Quantum (NISQ) devices, 
nonetheless, one cannot deny the importance of 
designing methods and algorithms that are realizable 
on quantum computers to span many-body Hilbert space\cite{RevModPhys.92.015003,PhysRevA.64.022319,Tilly_2022}.

Historically, the quantum phase estimation algorithm 
(QPEA), proposed by Abrams and Lloyd~\cite{PhysRevLett.79.2586,PhysRevLett.83.5162}, was the first
quantum algorithm towards the simulation
of state energies of a many-fermionic system. QPEA
relies on the construction of an input reference state 
that is projected onto the target eigenstate via a 
unitary operation~\cite{https://doi.org/10.1002/qua.27021,B804804E,doi:10.1126/science.1113479}. Although theoretically appealing, 
the long coherence time required 
for the such unitary evolution due to extremely deep 
and complex quantum circuits warrants the availability 
of sufficiently large number of fault-tolerant qubits 
which is still beyond experimental realization. The
variational quantum eigensolver\cite{Peruzzo_2014} provides a lucrative
alternative platform for the digital simulation of 
quantum many-body systems in NISQ devices. 

VQE has been realized experimentally in various 
quantum hardware architectures like the photonic 
quantum processors~\cite{Peruzzo_2014}, superconducting quantum
processors\cite{PhysRevX.8.011021} and trapped ion architectures\cite{PhysRevX.8.031022}. Significant
theoretical developments of VQE tailored for various
hardware platforms immediately followed its experimental
demonstration. VQE presumes the prior
knowledge of the many-body Hamiltonian matrix elements 
from the classically computed values and aims to 
finding the "best" variational approximation to the 
ground state wavefunction of the Hamiltonian starting 
from a 
trial wavefunction Ansatz. The trial wavefunction is 
constructed in terms of a series of tunable parameters
which provide its variational flexibility. The state
of the system is prepared through a parametrized 
quantum circuit, followed by computing the 
expectation value of the Hamiltonian through repeated 
measurements. Both these steps 
are carried out in quantum hardwares while the update 
of the parameters take place via some classical 
optimization in classical hardware. The attractive 
feature of VQE is that the state
preparation can be done via shorter quantum circuits 
compared to QPEA, making it a more desirable 
candidate for NISQ devices. Needless to say, the 
VQE results and the associated energy landscape often
depend critically on the choice of the parametrized 
ansatz~\cite{doi:10.1021/acs.jctc.0c00421,D1CP04318H,Tilly_2022}.

One of the most significant developments towards the 
practical realization of 
the unitary evolution in a quantum computer directly
originated from the unitary variant of coupled cluster 
theory (UCC)~\cite{Shen_2017,Romero_2019,doi:10.1063/1.5133059,doi:10.1063/1.5141835,D1CS00932J,Peruzzo_2014} 
with a given truncation in the rank of the 
cluster expansion. While the UCC ansatz with single and 
double excitation operators (UCCSD) provides a compact
wavefunction ansatz, they work best in the cases where
the molecular non-dynamic correlation is somewhat 
weak. Several efforts have been made to make the UCC
work for areas of strong correlation, either by increasing 
the excitation rank at the expense of higher implementation
overhead or by including generalized hamiltonian matrix 
elements in the wavefunction parametrization. The latter 
class of methods stems from Nakatsuji's density 
theorem\cite{PhysRevA.14.41} and Nooijen's conjecture\cite{PhysRevLett.84.2108} about 
the possibility of expressing the eigenstates of a 
Hamiltonian containing one and two body terms in terms of 
generalized excitation operators. While theoretically 
exact, a direct VQE implementation with all generalized\cite{Lee_2018}
excitation operators incurs high implementation cost
which is far beyond NISQ realization. The excitation
list can further be pruned by keeping only generalized
paired terms, giving rise to k-UpCCGSD ansatz\cite{Lee_2018} for which 
the circuit depth grows only linearly with the system size. 
In a different approach, the present authors proposed a partially 
disentangled form of the unitary in which the high rank 
excitations are implicitly folded in through nested 
commutators of two-body scattering operators with
effective hole-particle excitation rank of one and 
the standard cluster operators\cite{doi:10.1063/5.0114688}.

The goal for an expressive ansatz is to span the 
$N-$electron Hilbert space through a set of connected 
operators, ideally of rank $N$ at max. While keeping the
maximum rank of the operators (that enter the
parametrization) two for better NISQ realization, 
one must be able to express any $N-$body connected 
hole-particle excitation through low power cumulative
actions of different non-commuting operators\cite{doi:10.1063/1.4985916, doi:10.1021/acs.jctc.0c00736}. 
The action of any
generalized operator with effective excitation rank zero
on functions of the so-called \textit{primary excitation 
subspace} (spanned by the set of $n-$tuply excited
determinants; \textit{vide infra}) either leads 
to nilpotent solution, or results in redundant functions
within the primary excitation subspace. 
The two-body operator with effective excitation rank 
one (to be referred as scattering operator or 
scatterers, henceforth) can of course recursively 
lead to $(n+1), (n+2), (n+3)...)-$tuply excited 
functions to span the 
secondary excitation subspace. However, for such 
non-trivial action to exist, the specific scatterer 
(that leads to a determinant of the secondary 
excitation subspace by its action on a given 
primary excited determinant) must not commute with 
the cluster operator which generates the primary 
excitation function. As such, in this work, we pre-screen
the scatterers on-the-fly in terms of their 
(non-)commutativity with the cluster operators to come up 
with a dynamically optimal solution protocol: the COMmutativity
Pre-screened Automated Selection of Scatterers (COMPASS). 
COMPASS can be implemented in a parallel quantum 
architecture to choose the energetically most 
significant cluster operators to span the primary
excitation manifold and select the scatterers from 
an operator bath based on the commutativity criteria 
to span the secondary excitation manifold.
The entire ansatz may be constructed in a parallel
quantum environment and can be implemented with extremely
shallow quantum circuit.

Following the discussions on the genesis of COMPASS, we would
discuss its performance for the cases of molecular strong 
correlation in which conventional UCCSD fails to achieve 
desirable accuracy. In particular, we would study the
accuracy of COMPASS in handling molecular strong correlation, 
vis-a-vis the number of parameters needed. Furthermore, we would
analyse the accuracy of COMPASS in the case of noisy
simulations to demonstrate its expected performance in 
faulty quantum devices and would argue its suitability 
as a leading candidate for NISQ realization.

\section{Theory}
\subsection{Choice of the Operators Class and Motivation towards the Genesis of COMPASS:}
In order to design a compact parametrized ansatz, the
choice of the operators play the pivotal role. As discussed 
in the introduction, we choose to work with a set of one and 
two-body cluster operators along with a set of scatterers. 
The scatterers are two-body operators with effective hole-particle
excitation rank one and have one quasi-hole or quasi-particle
"destruction operator". Each element of the scatterers 
is equipped with an in-built projector such that it has
non-trivial action on only a selected set of primary 
excitation subspace determinants:
\begin{eqnarray}
S_{h} = \frac{1}{2} \sum_{amij} s^{am}_{ij} \{a^\dagger m^\dagger
ji\}|\cdot\cdot i \cdot\cdot j \cdot\cdot \cancel{m} \cdot\cdot
\cancel{a} \cdot\cdot \rangle \nonumber \\
\langle \cdot\cdot \cancel{a} \cdot\cdot 
\cancel{m} \cdot \cdot j \cdot\cdot i \cdot\cdot|
\\
S_{p} = \frac{1}{2} \sum_{abie} s^{ab}_{ie} \{a^\dagger b^\dagger 
ei\} |\cdot\cdot i \cdot\cdot e \cdot\cdot \cancel{a} \cdot\cdot 
\cancel{b} \cdot\cdot \rangle \nonumber \\ 
\langle \cdot\cdot \cancel{b} \cdot\cdot \cancel{a} 
\cdot\cdot e \cdot\cdot i \cdot\cdot |
\end{eqnarray}
Here, \textit{a, b, c, ...,} etc. denote the set of
unoccupied particle orbitals and \textit{i, j, k, ...,} 
etc. are the set of occupied hole orbitals with respect to the 
HF vacuum. Note that, \enquote*{\textit{m}} is a hole
state and \enquote*{\textit{e}} is a particle state 
with respect to HF vacuum and they together form a 
contractible set of orbitals (CSO).
The orbitals constituting CSO appear as the quasi-hole 
and quasi-particle destruction operators in $S$.
Thus, the operator $S$ that contains the label $m$ (or $e$) 
may be denoted as $S^m$ (or $S_e$). Similarly, the cluster operator 
$T$ with orbital label $m$ (or $e$) 
may be denoted as $T_m$ (or $S^e$). Note that for the cluster
operators, both $m$ ($e$) appears as quasi-hole (quasi-particle)
\textit{creation} operators whereas they appear as \textit{destruction} 
operators in $S$. As such, the contractions 
between $S$ and $T$ take place through them to simulate a
higher excitation rank operator. Each contraction 
between the scatterers and the cluster operators 
increases the effective hole-particle excitation rank 
by one, for example,
${\contraction{}{S}{}{T_2}
S T_2}\rightarrow T_3$, 
$\contraction{}{S}{}{S} \contraction[2ex]{}{S\;}{T_{2}}{} S\; S T_{2}
\rightarrow T_4$... .
The anti hermitian counterpart of the $S$ and $T$ operators 
($\sigma=S-S^{\dagger}; \tau=T-T^{\dagger}$) may be used
to construct an unitary evolution operator which can be 
implemented in quantum architecture. With the knowledge of 
the (non-)commutativity among various
$\sigma^m=S^m-{S^m}^{\dagger}$ and $\tau_m=T_m-{T_m}^{\dagger}$ 
(and among $\sigma_e=S_e-{S_e}^{\dagger}$ and $\tau^e=T^e-{T^e}^{\dagger}$), 
the effects of connected high rank excitations can be built 
through nested commutators like 
[$\sigma,\tau_2$] $\rightarrow 
\tau_3$; [[$\sigma,\tau_2$],$\tau_2$]$ \rightarrow 
\tau_4$... and so on. As such a partially disentangled 
(factorized) ansatz can be shown to include the infinite 
commutators even when a finite Trotter order~\cite{article} is used 
to approximate $e^\tau$ and/or $e^\sigma$.
\begin{equation}
    e^{\sigma}.e^{\tau}= e^{\sigma+\tau+[\sigma,\tau]+[[\sigma,\tau],\tau]+...}
\end{equation}
Note that the action of the unitary $e^{\sigma}$
on the $e^\tau |\Phi_{HF}\rangle$ is partially 
nilpotent due to
the inbuilt projector in the definition of $\sigma$. 
Expanding the entangled states $e^\tau|\Phi_{HF}\rangle$
in terms of constituent zero, one, two,... body 
excited determinants (which belong to the primary
excitation manifold), one may write
\begin{equation}
    e^{\sigma}\Big(e^{\tau}\ket{\Phi_{HF}}\Big)\rightarrow e^{\sigma}\Big(|\Phi_{HF}\rangle + \sum c_I|\Phi_I\rangle\Big)
\end{equation}
Here $\Phi_I$'s are the various determinants belonging 
to the primary excitation manifold
generated by $e^{\tau}\ket{\Phi_{HF}}$. The index $I$
would generically denote the composite hole-particle
indices associated with the excited determinants or with
the cluster operators. Note that each $\sigma$ operator 
is characterized by a quasi-hole/particle
destruction operator that belongs to CSO. The action 
of $\sigma$ is nontrivial only on certain set of primary subspace
determinants. For example, a $\sigma$ operator with 
quasi-hole destruction orbital '$m$' acts only on those 
excited primary subspace determinants where the 
occupancy of '$m$', $n_m$
is zero. For each $\sigma^m$, the occupancy of '$m$' 
may be used to divide the primary subspace determinants 
on which $\sigma^m$ acts into two sets.
\begin{equation} \sum_{I}^{N}c_{I}|\Phi_{I}\rangle=\sum_{J=1}^{N_1}c_{J}|\Phi_{J}(n_m=0)\rangle + \sum_{K=1}^{N_2}c_{K}|\Phi_{K}(n_m=1)\rangle
\end{equation}
with $N=N_1+N_2$ represents the total number of 
determinants generated by $e^\tau|\Phi_{HF}\rangle$ that span 
the primary subspace. 
Note that for a hole state '$m$', $e^{\sigma^m}(\sum_{J}c_{J}|\Phi_{J}(n_m=0)\rangle)$ has 
non-vanishing action while 
$e^{\sigma^m}(\sum_{K}c_{K}|\Phi_{K}(n_m=1)\rangle)$ is nilpotent. 
These determinants with $n_m=0$ are principally generated 
by $\tau_m$ and these cluster operators do not commute with $\sigma^m$.
Unfortunately, an unrestricted construction of the
entangled state $\sum_{I}c_{I}|\Phi_{I}(n_m=0\oplus n_m=1)\rangle)$ utilizes high number of parameters, 
resulting in a deep circuit. An exact similar analysis 
can be done when the destruction operator is a 
quasi-particle state that belongs to CSO. 

Instead, one may selectively generate only those 
primary subspace determinants by the cluster operators 
on which the scatterers have a non-trivial action. The 
COMPASS dynamically chooses the "best" set of cluster 
amplitudes in a factorized manner. Each of the primary 
subspace determinants, generated by the individual 
cluster amplitudes, are scanned for whether a non-commuting 
scatterer (with which the corresponding cluster operator share CSO) 
have significant effect or not. Where the commutativity
criteria is met, the scatterers are immediately allowed 
to act upon the entangled state. In the next section, 
we would present the genesis of COMPASS and discuss upon
how it directs to the construction of the most 
optimal ansatz.

\subsection{Development of COMPASS: an Automated Toolkit for Dynamic Ansatz Design}
\begin{figure*}[t]
    \centering   \includegraphics[width=18.2cm,height=12.6cm]{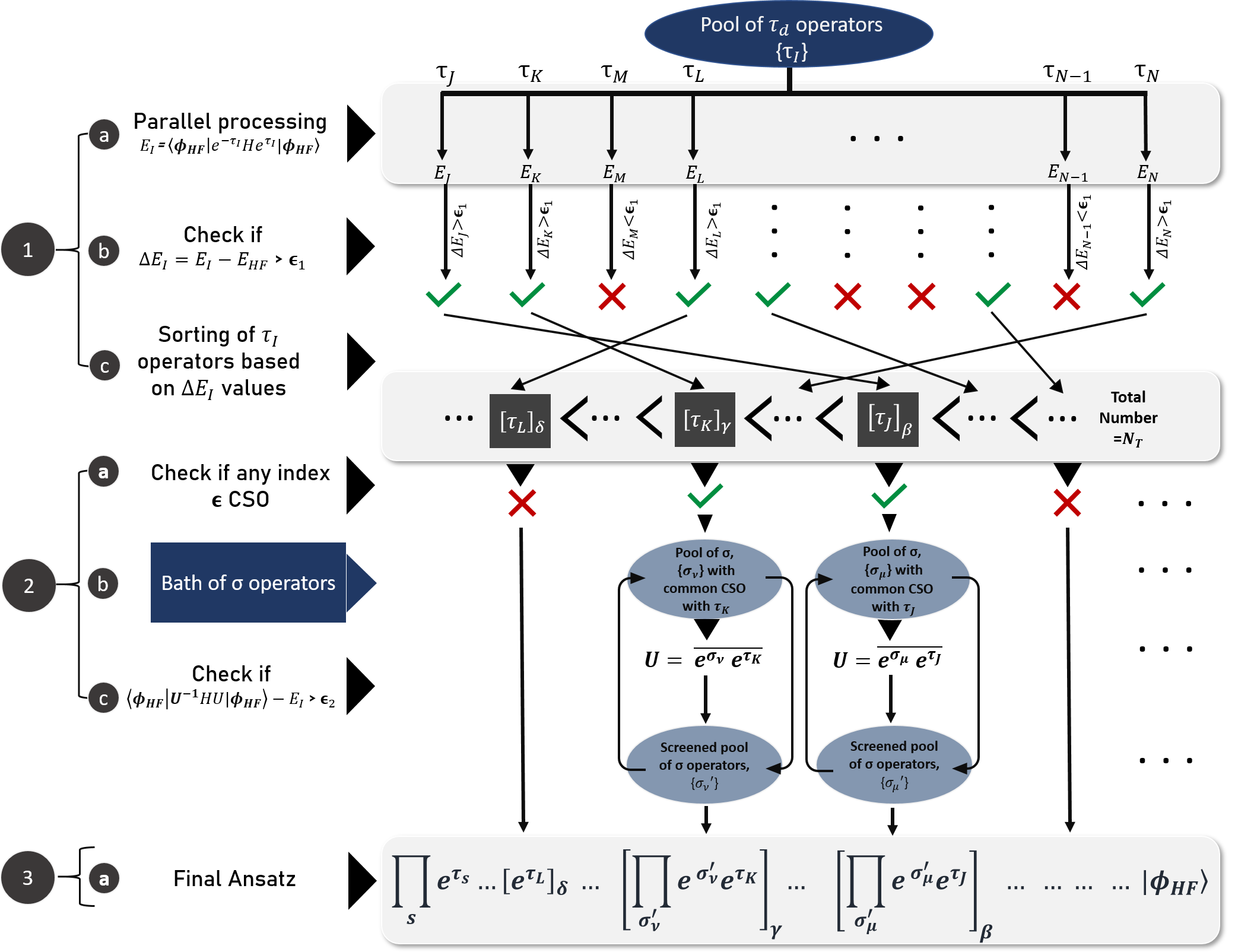}
    \caption{\textbf{Schematic representation of COMPASS protocol. In \textbf{step 1(a-c)}, each cluster operator $\tau$ is screened (potentially in a massively parallel quantum architecture) through a threshold criteria and kept in an operator block as per their descending contribution to correlation energy. Each such selected $\tau$ is checked if that contains any orbital(s) that belongs to CSO \textbf{(step 2(a))}. For each operator block with selected $\tau$ operator, appropriate non-commuting $\sigma$ operator(s) are fetched from the operator bath. The blocks are expanded by including the energetically dominant $\sigma$ operators (denoted in this figure as $\sigma^\prime$), one at a time, selected via the optimization of two-parameter energy functionals \textbf{(step 2(b))}. This step can again be performed in a parallel quantum architecture. The operator bath, in principle, may supply infinite number of $\sigma$ operators and is never drained out. The final ansatz is constructed by concatenating the various operator blocks, keeping the ordering unchanged \textbf{(step 3(a))}.}}
    \label{fig:Compass}
\end{figure*}
COMPASS relies on the choice of the "best" set of the 
cluster operators and selection of the appropriate 
scatterers that have significant contribution by their
action on certain entangled states. The whole ansatz 
is constructed dynamically, possibly in a parallel
quantum architecture and the resulting ansatz 
features as a disentangled product of various 
$e^{\sigma_\mu}$ and $e^{\tau_I}$ in an interwoven manner.
As mentioned before, the indices ${I, J, K, L,...}$ 
are used to denote the composite excitation labels
associated with $\tau$ whereas ${\mu, \nu, \lambda}$ 
would denote the composite orbital labels for $\sigma$.
COMPASS consists of three components: a parallel selection
of the "best" amplitudes and energy ordering, 
selection of the appropriate scatterers from an 
operator bath via CSO scanning, and the preparation 
of the final ansatz. A schematic figure of COMPASS
is presented in Fig \ref{fig:Compass} and the details 
of the same is discussed below.

\subsubsection{Choice of the "best" cluster operators and the ordering of their appearance:}
The choice of the "most important" cluster operators is
a key step to design a shallow depth quantum circuit in 
the NISQ devices~\cite{https://doi.org/10.48550/arxiv.2106.15210}. Towards this, we have considered only 
the double excitation operators (generically to be
denoted as $\tau_d$) while the single excitation
cluster operators (generically to be
denoted as $\tau_s$) will be handled separately. With 
the availability of multiple quantum devices, one may
parallelly optimize the various one parameter energy functional:
\begin{eqnarray}
\label{one_param_E}
E_{I} &=& \bra{\Psi(\theta_I)}\hat{H}\ket{\Psi(\theta_I)}, \nonumber \\ 
&=& \bra{\Psi_{HF}} e^{-\tau_I} H e^{\tau_I} \ket{\Psi_{HF}} \\ &\forall& I \in (1, n_o^2n_v^2) \nonumber
\end{eqnarray}
Each of these can be done through a one parameter circuit, 
with $\theta_I$ being the sole variational parameter. 
Here $\ket{\Psi(\theta_I)}= e^{\tau_{I}(\theta_I)} \ket{\Psi_{HF}}$. 
If sufficiently large number of quantum computers are available, each $E_I$ can be evaluated on different 
devices simultaneously. Otherwise, on a single 
quantum device one may repeat this multiple times and
take an average of $E_I$.
Only those $\tau_I$ operators will be 
kept in the cluster operator pool for which $\Delta E_I = | E_I - E_{HF}| > \epsilon_1$, where $\epsilon_1$ is
a predefined threshold and the rest of the cluster
operators are discarded.

With the $N_T$ number of cluster operators that pass
through the energy screening, we align them in a 
descending order of their contribution to correlation 
energy. Energetically most contributing cluster 
operator is allowed 
to act on the HF state directly, followed by the next 
and so on. This implies that for $N_T$ cluster operators
with $...\Delta E_J>...>\Delta E_K>...>\Delta E_L...$, $e^{\tau_{J}}$ 
is placed in operator block $\beta$, $e^{\tau_{K}}$ in 
operator block $\gamma$, ... and so on in a disentangled manner.
Note that $...\delta > ...>\gamma>...>\beta>...>1$
\begin{eqnarray}
    |\Psi\rangle &=& ...\Big[e^{\tau_L}\Big]_{\delta}... \Big[e^{\tau_K}\Big]_{\gamma}... \Big[e^{\tau_J}\Big]_{\beta}...|\Psi_{HF}\rangle \nonumber \\
    |\Psi\rangle &=& \prod_{\alpha=1}^{N_T} \Big[e^{\tau_I}\Big]_{\alpha} |\Psi_{HF}\rangle
\end{eqnarray}
where the quantities inside parenthesis denote the 
order of the operator blocks in which they act 
on $|\Psi_{HF}\rangle$.
This ordering will be maintained throughout. In general, if a given
$\tau_I$ is placed in operator block $[..]_\alpha$, we
would denote it as $[\tau_I]_\alpha$.
We mention that each cluster operator is allowed to 
appear only once irrespective of the block it is placed.
In the next subsection, 
we will dynamically expand the operator blocks by 
placing the scatterers appropriately 
based on the commonality of CSO labels shared 
between the cluster operators and the scatterers. 

\subsubsection{Selection of the Scatterers from Operator Bath}
For each operator $[\tau_I]_\alpha$ selected through 
the energy screening, one first checks if $\tau_I$ 
contains any orbital that belongs to CSO. For the 
cluster operators which do not contain any index of CSO,
the corresponding operator block is not expanded any
further. For all other cluster operators with one or more
orbitals that belongs to CSO, a pool of scatterers is 
created from the scatterer bath through commutativity
screening. This implies that the cluster operator and 
the scatterers should contain the same set of
contractible orbital(s). Thus for a given $[{\tau_{{m}_I}}]_\alpha$, 
the scatterer pool contains all the operators of the 
structure like $\sigma^m$. With $N_\alpha$ number of 
such scatterer selected in the pool (corresponding to the 
operator block $\alpha$ containing $\tau_I$), one optimizes the following two parameter energy functional:
\begin{eqnarray}
\label{two_params_E}
E_{I\mu} &=& \bra{\Psi(\theta_I,\theta_\mu)}\hat{H}\ket{\Psi(\theta_I,\theta_\mu)}, \nonumber \\ 
&=& \bra{\Psi_{HF}} \overline{e^{-\tau_I} e^{-\sigma_\mu}} H \overline{e^{\sigma_\mu} e^{\tau_I}} \ket{\Psi_{HF}} \\ 
&\forall& \mu \in (1, N_\alpha) \nonumber
\end{eqnarray}
Here the overline suggests the connected action of 
$\tau_I$ and the various 
$\sigma_\mu$'s that share \textit{at least} one 
common orbital index
that belongs to CSO. However, only those scatterers are 
screened for which $|E_{I\mu} - E_I | > \epsilon_2$,
where $\epsilon_2$ is a predefined threshold
whose value is usually taken to be order of magnitude
less than $\epsilon_1$. With the energy condition met, 
the operator block $[...]_\alpha$ is expanded by including 
appropriate $\sigma_\mu$'s through disentangled 
(factorized) unitary. 
\begin{eqnarray}
U_\alpha &=& \Big[ \prod_\mu \overline{e^{\sigma_\mu} e^{\tau_I}}\Big]_\alpha  
\label{eqemunu_ansatz}
\end{eqnarray}
There may, of course, be specific cases where the $\tau$ operator
in a given operator block does not contain any orbital of
CSO and in that case, as mentioned before, no
$\sigma$ gets attached to it. However, it still gets
placed according to its energy ordering.
Note that, for general cases, with each $\sigma_\mu$ 
being selected one-by-one for a given $\tau_I$,
the optimization in Eq. \ref{two_params_E} can be done with
shallow two-parameter circuit with $\theta_I,\theta_\mu$
as the two variational parameters. One may further note 
that the selection of scatterers for various
operator blocks are independent of each other and 
thus the evaluation of $E_{I\mu}$ in 
Eq. \ref{two_params_E} for various pairs
of $\tau_I$ and $\sigma_\mu$ can be 
done with multi-level parallelization. Furthermore, if 
sufficiently large number of quantum computers are 
available, one may evaluate the energy functional 
$E_{I\mu}$ in different devices simultaneously or
otherwise, one may perform it multiple times in a 
single quantum device and take an average value.
We further note that while a given $\tau_I$ appears
only once in the ansatz, a given $\sigma_\mu$ may be
attached to various $\tau_I$'s in their respective 
operator blocks and thus the $\sigma$ operators are
never drained out from the operator bath. 

\subsubsection{Weaving the Final Ansatz}
The final ansatz is constructed by concatenating the 
various operator blocks. As mentioned above, depending 
on the commonality of the contractible orbitals between 
the $\tau$ and $\sigma$ operators, several $\tau_I$ 
operator may attach same $\sigma$ operator to appear 
in their respective operator blocks.
Finally, all the one-body cluster operators, 
$\tau_s$, are placed at the end in lexical ordering.
\begin{equation}
    U = \prod_{s} e^{\tau_{s}}\prod_\alpha U_\alpha  
    \label{finalansatz}
\end{equation}
Once the full ansatz is constructed for a given molecule 
with a fixed nuclear arrangement, the parameters involved 
in the ansatz, Eq. \ref{finalansatz} are optimized using
the standard VQE hybrid quantum-classical framework till
all the parameters are converged.

One may note that the 
proposed ansatz may span the entire $N-$electron 
Hilbert space. However, in COMPASS each $\sigma$ 
operator is chosen and clubbed with one or more 
non-commuting $\tau$ operator(s) in a way that lowers the
correlation energy beyond a certain predefined 
threshold. The selected $\sigma$ operators are allowed to 
act immediately upon certain entangled states that 
are generated by the action of the associated 
non-commuting $\tau$ on the Hartree Fock determinant. 
Note that only those
$\sigma$ operators are kept in a given operator block 
(having a pre-selected cluster operator $\tau$) that 
can generate at least a triply excited secondary excitation
subspace function.

\section{Results:}
COMPASS has been implemented with an interface to qiskit-nature~\cite{Qiskit}
which imports the one and two-electron integrals from 
PySCF~\cite{https://doi.org/10.1002/wcms.1340}. All the calculations performed in this study 
employed STO-3G basis set~\cite{doi:10.1063/1.1672392} with a direct spinorbital to
qubit mapping. The Jordan-Wigner transformation~\cite{doi:10.1063/1.4768229} was 
employed to encode second quantized operators to qubit
operators. For all our calculations, we chose the L-BFGS-B~\cite{morales2002numerical,doi:10.1137/0916069}
optimizer for the classical optimization. Also, we had 
initialized the qubits to the Hartree-Fock reference state
and each parameters of the ansatz was initialized to
the optimized values obtained from the minimization of energy functional of  Eq.\ref{two_params_E} (and Eq. \ref{one_param_E}).

While the $\sigma$ operators generically used so far 
towards the development have all possible spin- and spatially-unrestricted 
terms in it, in the actual implementation, the list is further 
significantly pruned by keeping only a few specific kinds of the same. 
Two different cases are considered which are
different from each other in the choice of the $\sigma$
operator taken in the operator bath. In particular, we 
have worked with the (a) opposite spin (OP) sector 
of $\sigma$ and (b) the partially paired (PP) sector 
of $\sigma$. 

The OP sector incorporates the specific low-spin channel 
of $\sigma$. This implies that the spins in the
excitation vertex and the scattering vertex of are 
different. 
\begin{eqnarray}\label{S_no_same_spin}
\sigma_{OP} \in \{{(\sigma_h)}_{i_{\alpha}j_{\beta}}^{a_{\alpha}u_{\beta}},
{(\sigma_h)}_{i_{\beta}j_{\alpha}}^{a_{\beta}u_{\alpha}};
{(\sigma_p)}_{i_{\alpha}v_{\beta}}^{a_{\alpha}b_{\beta}},
{(\sigma_p)}_{i_{\beta}v_{\alpha}}^{a_{\beta}b_{\alpha}}
\}
\end{eqnarray}  
$u$ and $v$ are the spatial active hole and virtual
orbitals that form the CSO, and $\alpha$ and $\beta$ denote the spin-up and spin-down electrons, respectively.

The PP sector of $\sigma$ includes (i) only those 
quasi-hole creation operators which originate from 
the same spatial orbitals (for $\sigma_h$) and (ii) 
only those quasi-particle creation 
operators which share the same spatial orbitals 
(for $\sigma_p$). 
\begin{eqnarray}
\sigma_{PP} \in 
\{{(\sigma_h)}_{i_{\alpha}i_{\beta}}^{a_{\alpha}u_{\beta}},
{(\sigma_h)}_{i_{\beta}i_{\alpha}}^{a_{\beta}u_{\alpha}};
{(\sigma_p)}_{i_{\alpha}v_{\beta}}^{a_{\alpha}a_{\beta}},
{(\sigma_p)}_{i_{\beta}v_{\alpha}}^{a_{\beta}a_{\alpha}}
\}
\end{eqnarray}
Irrespective of the choice of the sector of $\sigma$,
the orbitals constituting CSO are restricted only to those
that span the chemically active region.

\begin{figure*}[t]
    \centering
    \includegraphics[width= 18cm, height=12cm]{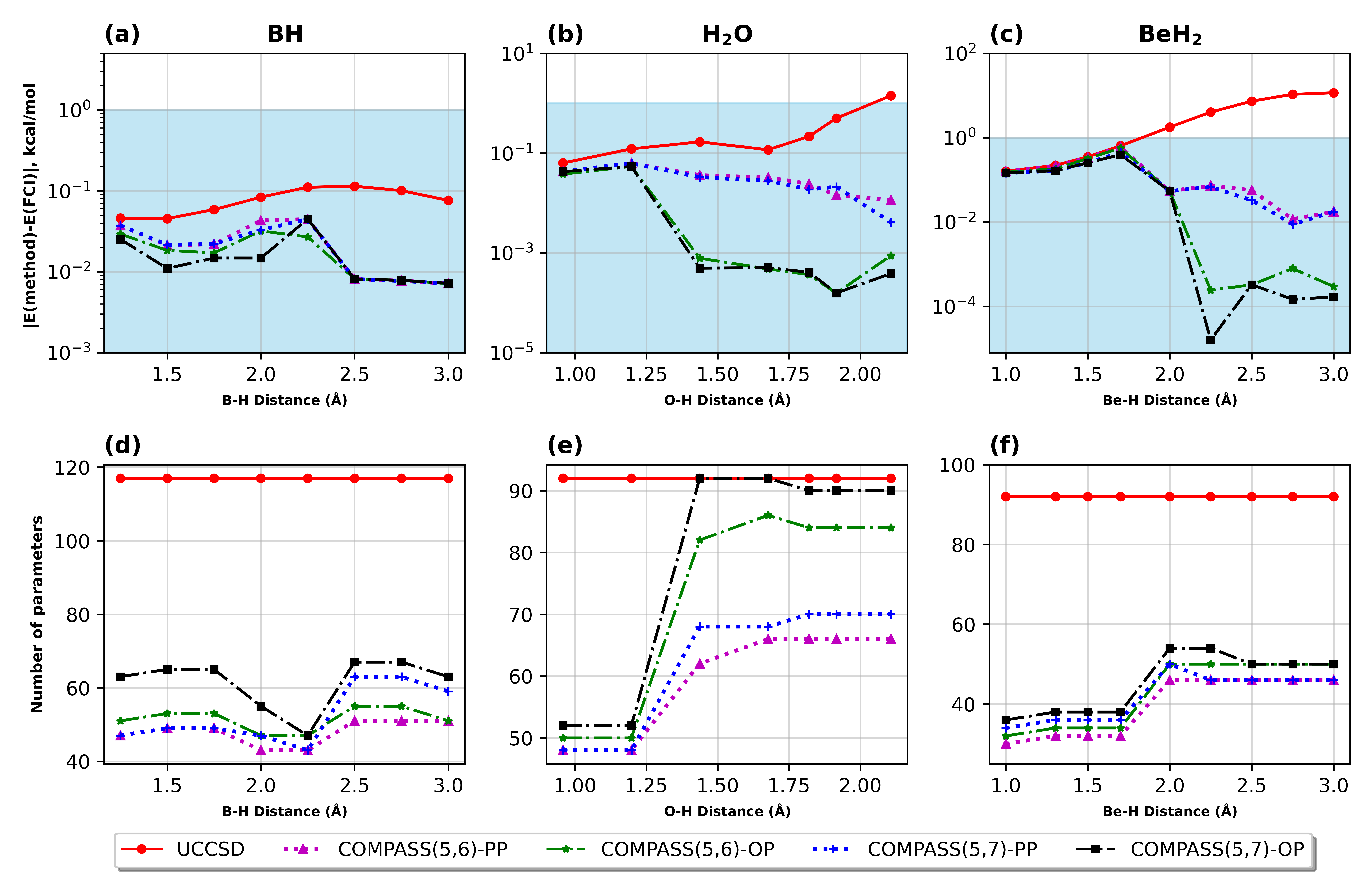}
    \caption{\textbf{Accuracy as a function of the bond length parameter for various versions of COMPASS with respect to FCI: (a) $BH$, (b) $H_2O$, and (c) linear $BeH_2$. The region shaded in pale blue indicates chemical accuracy in kcal/mol. (d), (e) and (f) estimate parameter counts for $BH$, $H_2O$ and linear $BeH_2$, respectively, along the potential energy profile.}}
    \label{fig:results}
\end{figure*}

\begin{figure*}[t]
    \centering
    \includegraphics[width= 18cm, height=3cm]{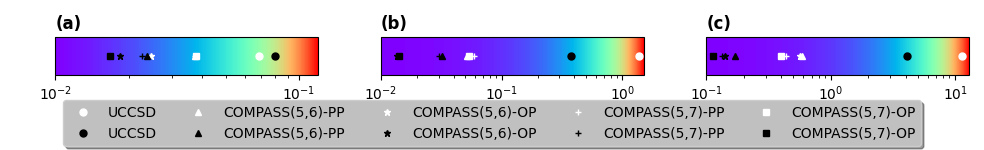}
    \caption{\textbf{NPE (white symbols) and average error (black symbols) (in kcal/mol) across the potential energy profile for (a) $BH$, (b) $H_2O$, and (c) linear $BeH_2$ models. For $H_2O$, the NPE shows two orders of magnitude improvement over the UCCSD while for the linear $BeH_2$, all variants of COMPASS have several orders of magnitude improvement in NPE, estimating a uniform description of many-body correlation effects over the entire energy profile. The color-coded horizontal axis denotes the order of NPE and average errors.}}
    \label{fig:npe}
\end{figure*}

As discussed previously, COMPASS uses two threshold 
parameters ($\epsilon_1$ and $\epsilon_2$) which ensure
to include the most dominant cluster amplitudes and the 
scatterers in the final ansatz. As such, COMPASS 
calculations with thresholds of $\epsilon_1$ and 
$\epsilon_2$ would be denoted as 
COMPASS($-log(\epsilon_1),-log(\epsilon_2)$). 

\subsection{Accuracy vs Parameter Count over the Potential Energy Profile of Strongly Correlated Molecules:}

Towards the study of 
the accuracy of COMPASS, three difficult test cases have
been identified: the potential energy surface for the 
the stretching of $BH$ single 
bond, the symmetric simultaneous single bond stretching 
of $H_2O$ and linear $BeH_2$. 
In all cases, the results are compared against 
UCCSD. In all our calculations, all the 
amplitudes corresponding to the spin-complemented 
operators are treated as independent. 

Stretching of the single bond in $BH$ is one of the most
difficult test cases for assessing the accuracy of any
quantum many-body theory. The system shows the signature 
of strong molecular correlation due to the interplay of
the ground and excited roots when the $B--H$ bond is 
stretched. We model its dissociation profile with COMPASS
and UCCSD. BH consists of 6 electrons in 12 spinorbitals
with a total Hilbert space dimension of 4096. The 
UCCSD energy profile, when 
plotted as a function of $B--H$ distance shows energy 
error $\sim 1$ kcal/mol with respect to the
classically exact Full Configuration Interaction 
(FCI) method throughout various molecular
arrangements. COMPASS with both the
variants, OP and PP, outperform UCCSD throughout the 
energy surface and particularly when 
the molecule is sufficiently stretched, it shows
improvement by an order of magnitude. Interestingly
enough, COMPASS(5,7) with OP variant takes about 67
parameters at bond length of 2.5 and 2.75\AA, which is the
highest (compared to other nuclear arrangements) in terms 
of the number of parameters required for COMPASS, but
still this is an order of magnitude less than 
the 117 parameters (without combining the 
spin-complementary excitations) taken up by UCCSD. The 
accuracy throughout the potential energy profile is also
illustratively evident from the non-parallelity error (NPE) \
and the average error for various schemes and is shown in 
Fig. \ref{fig:npe}(a).

\begin{figure*}[t]
    \centering
    \includegraphics[width= 18cm, height=6cm]{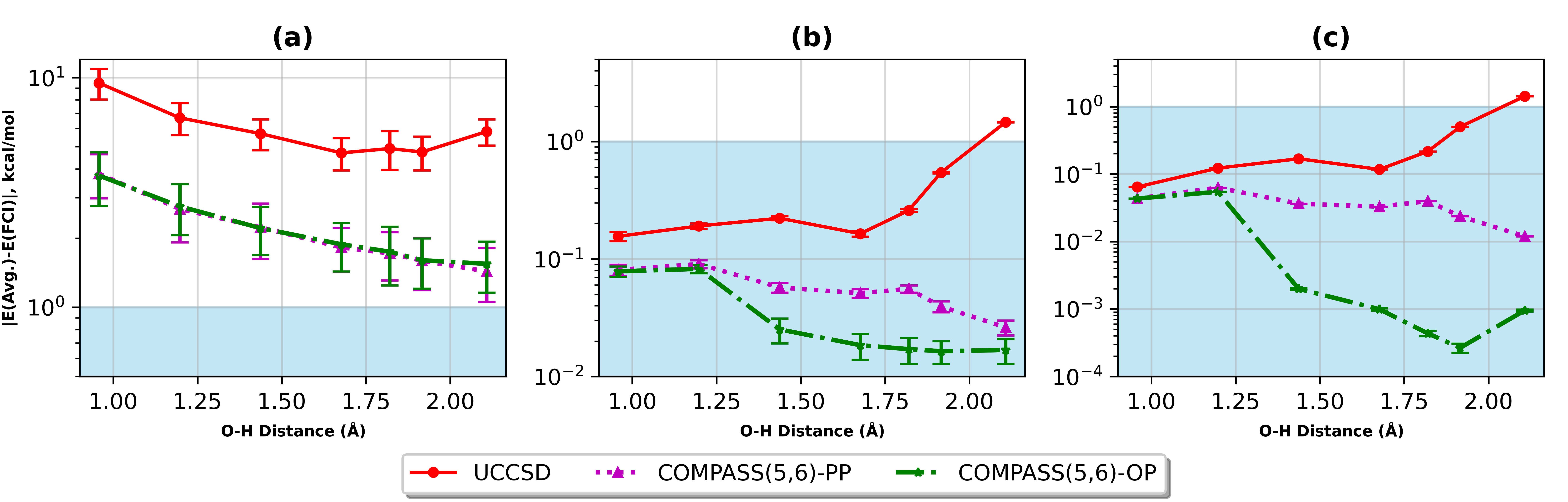}
    \caption{\textbf{Accuracy as a function of bond length parameters for $H_{2}O$ ($H-O-H=104.4776^{\circ}$) under gaussian noise model characterized by standard deviation $SD$ : (a) $SD=10^{-2}$, (b) $SD=10^{-3}$, (c) $SD=10^{-4}$. For both (b) and (c) the average energy predicted by COMPASS are well within chemical accuracy. For all the variants, COMPASS predicts energy which are at least an order of magnitude accurate than UCCSD across the energy profile; demonstrating its resilience to noisy environment. The vertical lines across each point suggest the standard deviation of the observed energy values with 100 independent samples.}}
    \label{fig:noise}
\end{figure*}

$H_2O$ in STO-3G basis, with one core spatial orbital
frozen, renders to be a system with 8 electrons in 12
spinorbitals. In Fig. \ref{fig:results}b, we have plotted the
energy error with respect to FCI method as a function
of the bond length parameter keeping the $H-O-H$ angle 
fixed at $104.4776^{\circ}$. While the UCCSD ansatz 
requires 92 parameters (without combining the spin-
complementary excitations), it fails to achieve energy
error within chemical accuracy when the bonds are
sufficiently stretched. Dramatically enough, all the 
variants of COMPASS with different threshold parameters
significantly outperforms UCCSD and resulting COMPASS 
energy values are just a few micro-Hartree away
from FCI throughout the energy surface. Even more
impressive is the significantly fewer number of 
parameters selected by most of the COMPASS variants 
to achieve this level of accuracy throughout the potential 
energy profile: COMPASS(5,6) with the PP variant 
barely needs about 66 parameters, resulting in a much
compact form of the wavefunction ansatz. Beyond 1.4 \AA, 
OP variants require higher number of parameters than
the PP counterpart; however, this also results in 
at least an order of magnitude improvement in accuracy. 
The overall superiority of COMPASS in comparison to UCCSD
is again measured in terms of NPE and the average error and
is deomnstrated in Fig. \ref{fig:npe}(b).

The efficacy of COMPASS is much more pronounced in 
strongly correlated systems like a linear $H-Be-H$ model
when we simultaneously stretch the $Be-H$ bonds in a 
symmetric manner. Due to the strong electronic correlation
on symmetric stretching of the bonds, the UCCSD behaves
somewhat poorly and beyond $R_{Be-H}=1.75$\AA, its 
energy error with respect to FCI goes beyond 1 kcal/mol.
The error further increases on further stretching the 
bonds, marking a clear signature of strong correlation. 
COMPASS with all the variants show remarkable improvement 
over UCCSD results with at least one order of magnitude 
in the region of short $R_{Be-H}$ length and by several
orders of magnitude when the bonds are further stretched.
This comes at a remarkable reduction in the number of 
parameter count over UCCSD: in the worst case scenario,
COMPASS(5,7) with OP variant utilises about 54 parameters
around $R_{Be-H}=2.00-2.25$\AA, while UCCSD requires 92
parameters. In fact, the OP variant shows several order of 
magnitude improvement in accuracy over the PP variant, particularly
in regions beyond 2 \AA. However, for two different sets of 
$\epsilon_1$ and $\epsilon_2$, both COMPASS-OP and COMPASS-PP
require similar number of parameters throughout the potential energy 
profile. The extremely high accuracy over the potential energy
profile is again measured in terms of the NPE and average error
(Fig. \ref{fig:npe}(c)) and COMPASS variants show clear two orders
of magnitude improvement of these metrics over UCCSD. 
The accuracy and parameter count in linear $BeH_2$ model
clearly demonstrate the performance of COMPASS in 
compactifying the ansatz for the simulation of strong 
electronic correlation, making it more suitable 
for NISQ realization.


\subsection{Simulation with Gaussian Noise Model:}
The simulation so far has presumed a noiseless 
implementation in an ideal quantum environment. However,
the NISQ devices are not fault-tolerant and as such in
practical scenario, one needs to account for 
the uncertainties due to imperfect implementation. 
Towards this, for each set of optimal parameters generated
in an ideal quantum setting, we randomly produce 100 
samples of noisy parameters derived via a Gaussian 
distribution model~\cite{2021npjQI...7..155M,PRXQuantum.2.030301}. For each parameter, the distribution
is centered at the optimal parameter value and the noisy
parameters ($\theta_m$) are generated as:
\begin{equation}\label{Gaussian}
\theta_{m}=exp\Big(-\frac{(\theta-\theta_{n})^2}{2 (SD)^{2}}\Big)
\end{equation}
Here $\theta_{n}$ is the optimal value of the parameters
and $SD$ is the standard deviation. Three different 
cases are considered with $SD=10^{-2}$, 
$SD=10^{-3}$ and $SD=10^{-4}$. With each of the 
100 samples taken into account for various $SD$, we
estimate the average energy. 

To estimate the performance of COMPASS with imprecise
implementation of the parameters, we choose to work with
$H_2O$ with the same set of geometrical parameters as 
in the case discussed earlier. We took a conservative
COMPASS parameters $\epsilon_1=10^{-5}, \epsilon_2=10^{-6}$.
In all various noise strengths, the different variants 
of COMPASS seem to be more resilient than UCCSD. With 
low noise strength ($SD=10^{-4}, SD=10^{-3}$), in the 
regions where the bonds are sufficiently stretched and 
the strong correlation dominates, COMPASS variants are
clearly a few orders of magnitude more accurate than 
the UCCSD counterpart (Fig. \ref{fig:noise}b, c). 
Only in the case of reasonably strong noise with $SD=10^{-2}$, 
COMPASS fails to achieve chemical accuracy though it is 
about an order of magnitude more accurate than UCCSD. 
One may note that in cases of such noisy simulations, the
optimal parameters are of the same order of magnitude
as the noise itself, and this may not potentially cause a
hindrance to its realistic implementation.

\section{Discussion and Future Outlook:}
Being a dynamic structured ansatz, it is difficult to 
comment on the resource requirements and the computational
cost of COMPASS. The measure for the resource requirement 
for a NISQ device is often taken to be the circuit depth 
and the number of measurements. While the number of 
measurement is an important parameter that determines 
the overall time to the solution, for a NISQ realization, 
the circuit depth may be taken as the most important
metric. One may note that the latter determines the 
feasibility of implementation of the algorithm in 
devices with short coherence time.   

The parent VQE algorithm suffers from the 
limitation of large number of measurements required 
to reach the solution. The COMPASS algorithm is likely
to suffer from the same drawback, if not slightly more. 
This mainly is due to the (a) the ansatz construction
(step 1,2, Fig. \ref{fig:Compass}) and (b) optimization
of the final ansatz. One may note that the ansatz
construction only requires several one and 
two-parameter energy functional optimization. Each 
of such optimization require very few number of iterations
and hence they involve few measurements. Most 
importantly, each of these energy optimizations 
are independent and thus can be performed in parallel 
quantum architecture. Thus, with the development of 
hardware capabilities, the requirement of increased 
number of shot count during the ansatz construction is
likely to pose limited challenges. However, the amplitude 
optimization towards the final state preparation may take
somewhat more number of steps (than standard UCCSD) 
due to the generalized operators present in the ansatz, 
resulting in a slower convergence. This is principally
attributed to the classical optimizer and one may adopt
various strategies to accelerate. The development of 
the fastest convergence algorithm would be studied 
in near future as it requires several aspects to take 
care of.  

We must point out that COMPASS is conceptually entirely 
different than the sequential growth ansatz like ADAPT-VQE~\cite{Grimsley_2019} 
where the operators are chosen based on the energy gradients
with respect to individual parameters from an operator pool.
COMPASS, on the other hand, is a dynamically expandable
ansatz construction protocol where no such energy gradient
needs to be computed to select the operators; rather, one
selects the "best" set of parameters through energy optimization
of one and two-parameter energy functionals in parallel
quantum architecture.
The tailoring protocol automatically ensures to 
include the most significant cluster amplitudes according 
to energetically optimal ordering. Appropriate scatterers
are immediately allowed to act upon once an entangled 
state is generated through the action of cluster operators 
on the Hartree Fock determinant. This has
shown to create a significantly compact and accurate 
wavefunction ansatz while excitations of arbitrarily high
rank can be captured through implicit commutativity 
between various class of operators. The COMPASS 
protocol tailors the ansatz depending on the system 
and its associated electronic complexity under 
consideration, while keeping the parameter count 
to the minimum. The compact
representation of the wavefunction makes it a desirable
candidate towards its realization on a NISQ platform.

One major advantage of COMPASS is its ability to 
build the wavefunction ansatz in a parallel architecture,
making the ansatz construction significantly less
impacted by the noisy environment. The accuracy is 
largely controlled by the two tuneable parameters. Even
with somewhat conservative choices of these parameters, 
the numerical results with various complex molecular
systems amply demonstrate its consistent superiority 
over UCCSD in terms of accuracy, gate count and the ease
of implementation. With high degree of quantum parallelism 
towards the preparation of the ansatz and its extremely 
low execution gate depth towards the state preparation
through the parametrized circuit, COMPASS would be an
extremely desirable candidate for digital molecular
simulations in near-term quantum devices.

\section{Acknowledgement:}
DM thanks Prime Minister's Research Fellowship (PMRF), Government of India for his research fellowship. SH thanks Council of Scientific and Industrial Research (CSIR), Government of India, and DH thanks Industrial Research and Consultancy Center (IRCC), IIT Bombay for their research fellowships.

\bibliography{literature}

\end{document}